# Counting and metrology of distributed atomic clocks using metropolitan fiber

**Jialiang Wang,** [1,2] **Deling Zeng,** [3,*] **Youzhen Gui,** [1,2] **Lian Dong,** [3] **Rong Wei,** [1,2,*]

[1] Key Laboratory for Quantum Optics, Shanghai Institute of Optics and Fine Mechanics, Chinese Academy of Sciences, Shanghai 201800, China

[2] Center of Materials Science and Optoelectronics Engineering, University of Chinese Academy of Sciences, Beijing 100049, China

[3] Shanghai Institute of Measurement and Testing Technology, Shanghai 201203, China

*Corresponding author: *weirong@siom.ac.cn*; *zengdl@simt.com.cn*

**abstract**

**We demonstrate a distributed atomic clocks network between Shanghai Institute of Optics and fine Mechanics (SIOM) and Shanghai Institute of Measurement and Test (SIMT). The frequency signals from three different clocks transfer in one fiber link and four clocks can have comparison in two different labs. By comparing the results of the comparison between the two labs, it was found that the consistency of the frequency signal is on the order of lower than 1E-15. And we also achieve consistency between two locations at the E-15 level of frequency reporting. This scheme can achieve distributed time counting and frequency dissemination of remote atomic clocks, which is a new exploration of the future time keeping laboratory mode.**

## Introduction.

In the past twenty years, frequency dissemination has been studied extensively and deeply because of its increasingly demand in many different applications, and the current optical frequency transfer with fiber link can achieve the stability of E-21. With the development of quantum frequency standards such as hydrogen maser and fountain clock, the short-term stability of radio frequency has reached 5E-14@1s and the most advanced optical frequency clock such as lattice clock and ion clock can have achieved stabilities better than E-18. The tidal effect and millimeter-level gravitational redshift can be observed through the optical clocks. At present, most of the high performance atomic clocks are placed outside of the time keeping laboratories, and many other users have also put forward application requirements for the highest precision frequency signals.

The field of atomic frequency scaling has made rapid progress in recent years, the existing clock time-keeping group mainly consists of hydrogen masers and fountain clocks, whose stability has reached the E-15magnitude or higher[1,2]. with optical frequency scaling reaching long-term stability and uncertainty on the order of E-19 and fiberoptic links enabling frequency signaling on the order of E-21. Frequency variations due to tidal effects, gravitational redshifts on the millimeter scale, have been seen through optical frequency scales. Currently, a large number of high performance atomic clocks are placed outside of timekeeping laboratories, while many laboratories are demanding the highest precision frequency signals available for their applications. This requires the generation and replication of time-frequency signals to be realized between different laboratories, which is what makes remote generation and realization of time-frequency signals the future direction of development in the field of time-frequency, as well as the future direction of technology development for generating TAI.

The existing clock-keeping group mainly consists of hydrogen clocks and fountain clocks, whose stability has reached E-15 magnitude or higher[1,2]. In order to realize the clock group long-distance clock comparison, need to use fiber optic time-frequency transfer technology to achieve a hundred kilometers or even longer distance atomic clock signal distribution[3,4], and finally realize the long-distance cooperative operation of the clock group[5-9], such as fountain clocks remote harnessing hydrogen clocks, remote time scale based on fiber network, etc. These techniques collect the dispersed clock signals to realize the common timekeeping of distributed clocks. This method concentrates the time and frequency signals in a laboratory, generates frequency standards and time standards and the transmits the signals to the user side, which is a 'decentralized-centralized-decentralized' method.

In this paper, we propose a scheme to distribute the time and frequency signals of different atomic clocks in two locations over long distances through the same fiber link to realize the frequency signals of all clock groups at each node to realize a distributed time and frequency laboratory, where each node in the laboratory system can get the frequency signals of all clocks, and the time and frequency signals between each node keep the same index. It has several advantages, first, so that a large number of clocks involved in the time keeping laboratory to generate time and frequency signal at the same time, the timekeeping capacity is greatly improved. Second, the laboratory is distributed in various locations, reducing the common mode noise of the clock caused by the environment, which is conducive to improving the uncertainty of the timekeeping; third, in each distributed timekeeping laboratory to generate high-performance time and frequency signals, more conducive to the user: fourth, the distributed timekeeping laboratory has better robustness.

A necessary condition for a distributed time-keeping laboratory is that each laboratory can share the clock signals of other laboratories, and compare them with the time scales distributed laboratories. This requires that the inter-laboratory time-frequency transfer link can transmit all the clock signals. This study mainly accomplished the pre-demonstration work of the above technology on the laboratory: we conducted the bi-directional frequency signal transfer and comparison between two laboratories at a distance of 40km and a fiber length of 80km, and compared the clock signals of the two places. The experimental results show that the consistency of the frequency comparison signal between the two places is which can realize the generation of the standard frequency signal in real time at the two places, and using this clock signal, we established a clock group of 5 clocks in the two laboratories

## Experiments and results

Our experimental link is shown in Figure 1(a). We establish a fiber optic time-frequency link between SIOM and SIMT based on 80km commercial communication fiber, which enables two time-frequency signals to pass from SIOM to SIMT, and one time-frequency signal to pass from SIMT to SOM. The time-frequency transfer system is shown in Figure 1(b), the100MHz hydrogen clock signal from SIOM is divided into two ways, one of which is modulated on Laser1, and the other is modulated on Laser2 after 10 times the frequency to 1GHz. The multiplied 1GHz signal is passed to SIMT through 80km field link and modulated on Laser3 and returned to the original path and identified with the reference source of SIOM to get the error signal and manipulate the fiber delay line to stabilize the link time delay. When the link is stabilized. the CH.3 modulated on Laser4 and CH.1 modulated on Laser1 can also be passed through the fiber optic link stably. SIMT has two hydrogen clocks, SIOM has two hydrogen clocks (VCH 1003A and iMaser 3000, respectively) and ones $^{87}$Rb fountain clock. The iMaser 3000 hydrogen clock of SIOM is placed in another laboratory at a distance of about 30m, connected by 1 100 m long from a fiber optic link, and the structure and layout of the experimental setup is shown in Figure 2. The purpose of this study is to achieve continuous frequency signal transmission between laboratories. The clock signals from each node laboratory are made to be concentrated in the timekeeping laboratory to generate distributed atomic time scales, and then the atomic time scales are transmitted to each node laboratory. The work in this paper focuses on completing the key techniques of this study, completing the sharing of clock signals between two distributed labs, and evaluating the performance of the transmission link. For simplicity, we use CH.1 and CH.2 to mark the two hydrogen clocks of SIMT, and CH.3, CH.4 and CH.5 to mark the two hydrogen clocks and one fountain clock of SIOM, as shown in Figure 2. L1 and L2 are used to mark the fiber optic link from SIOM to SIMT, L3 is used to mark the fiber optic link from SIMT to SIOM, and L4 is used to mark the fiber optic link inside SIOM.

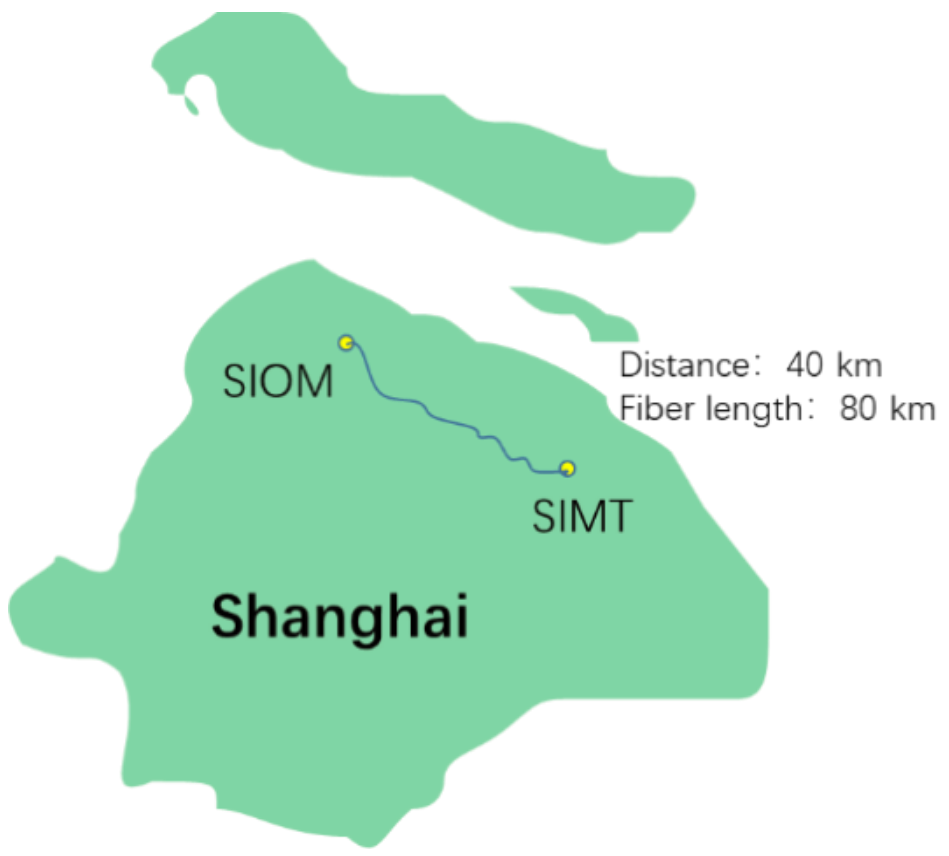

(a)

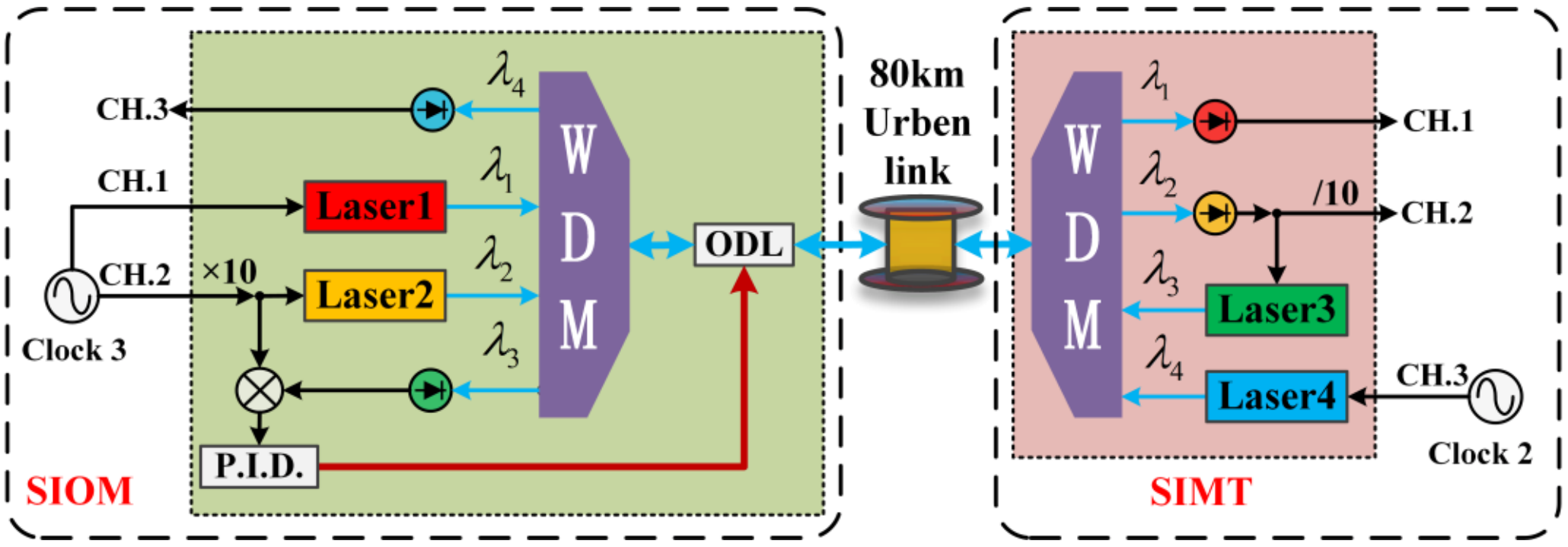

(b)

**Fig.1 (a) Link diagram of Shanghai distributed time-frequency laboratory. (b) Experiment setup diagram**

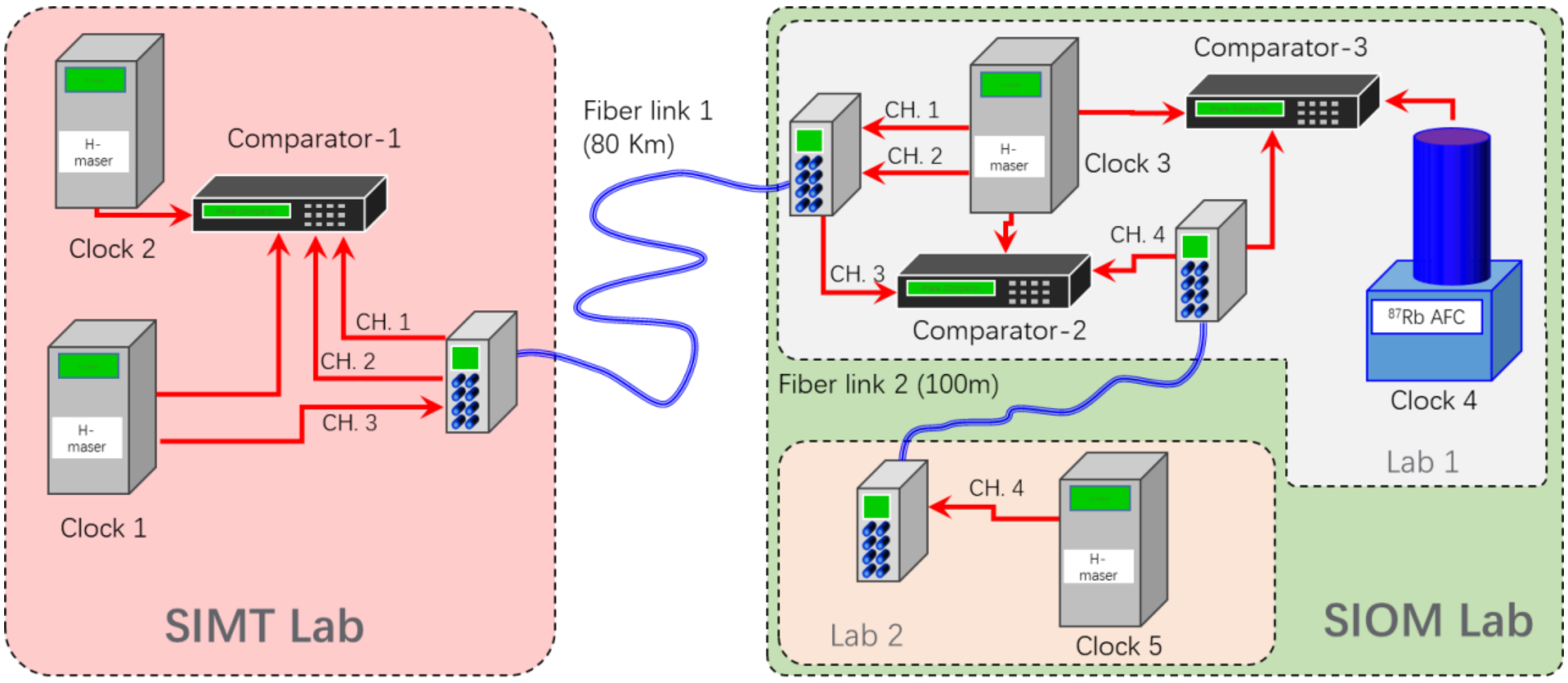

**Fig.2 Structure and layout of the distributed lab**

(To make some introduction to the multi-signal time-frequency transfer of single fiber, using a single channel monitoring feedback, multi-channel two-way simultaneous transmission method, the future is expected to further reduce the noise of the fiber optic link through the signal back transmission approach), the future is expected to achieve multi-clock frequency, time signal bi-directional transfer, this study mainly verified its performance. Therefore, we focused on the results of clock Cl and C3 with L1, L2, and L3 transfer signals, in SIOM and SIMT comparison, as a control, also tested the comparison results of clock C3 and C5 in two phase comparison instruments. We use $y_i(\tau) = (\nu_i - \nu_0)/\nu_0$ denotes the time from the start of the comparison experiment from(i time toil The relative frequency deviations from the beginning of the comparison experiment to the beginning of the comparison experiment. The comparison leads to the following

$$y_{\mathrm{SIMT1},i}(\tau) = \left[y_{\mathrm{C3},i}(\tau) + y_{L1,i}(\tau)\right] - y_{\mathrm{C1},i}(\tau) \qquad (1\text{-}1)$$
$$y_{\mathrm{SIMT2},i}(\tau) = \left[y_{\mathrm{C3},i}(\tau) + y_{L2,i}(\tau)\right] - y_{\mathrm{C1},i}(\tau) \qquad (1\text{-}2)$$
$$y_{\mathrm{SIOM},i}(\tau) = \left[y_{\mathrm{C1},i}(\tau) + y_{L3,i}(\tau)\right] - y_{\mathrm{C3},i}(\tau) \qquad (1\text{-}3)$$

Here $y_{\mathrm{C1}}$ and $y_{\mathrm{C3}}$ denote the frequency deviation of clocks C1 and C3. $y_{L1}$ and $y_{L2}$ and $y_{L3}$ denote the additional frequency deviation caused by the 3 links, and since all 3 links are realized through a single fiber, the $y_{L1}$ and $y_{L2}$ and $y_{L3}$ contains the common mode noise of the fiber optic link, and the difference between the 3 signals is caused by their use of different digital-to-analog conversion modules, different microwave transmission signals, different bar ratio phase meters, etc. From equation (1), if two of the three formulas are summed or differenced, the deviation between the two links can be obtained, but the actual situation is not so simple, because the comparison experiment requires that the two clock signals must be synchronized, which requires that the three comparison experiments start at the same time, the actual situation can not meet this requirement. We will discuss the results of the comparison experiments first, and then discuss the problem of signal synchronization. This place plus the graph of stability and relative deviation shows that high uncertainty can be achieved in a very short time with consistent frequency deviation using links.

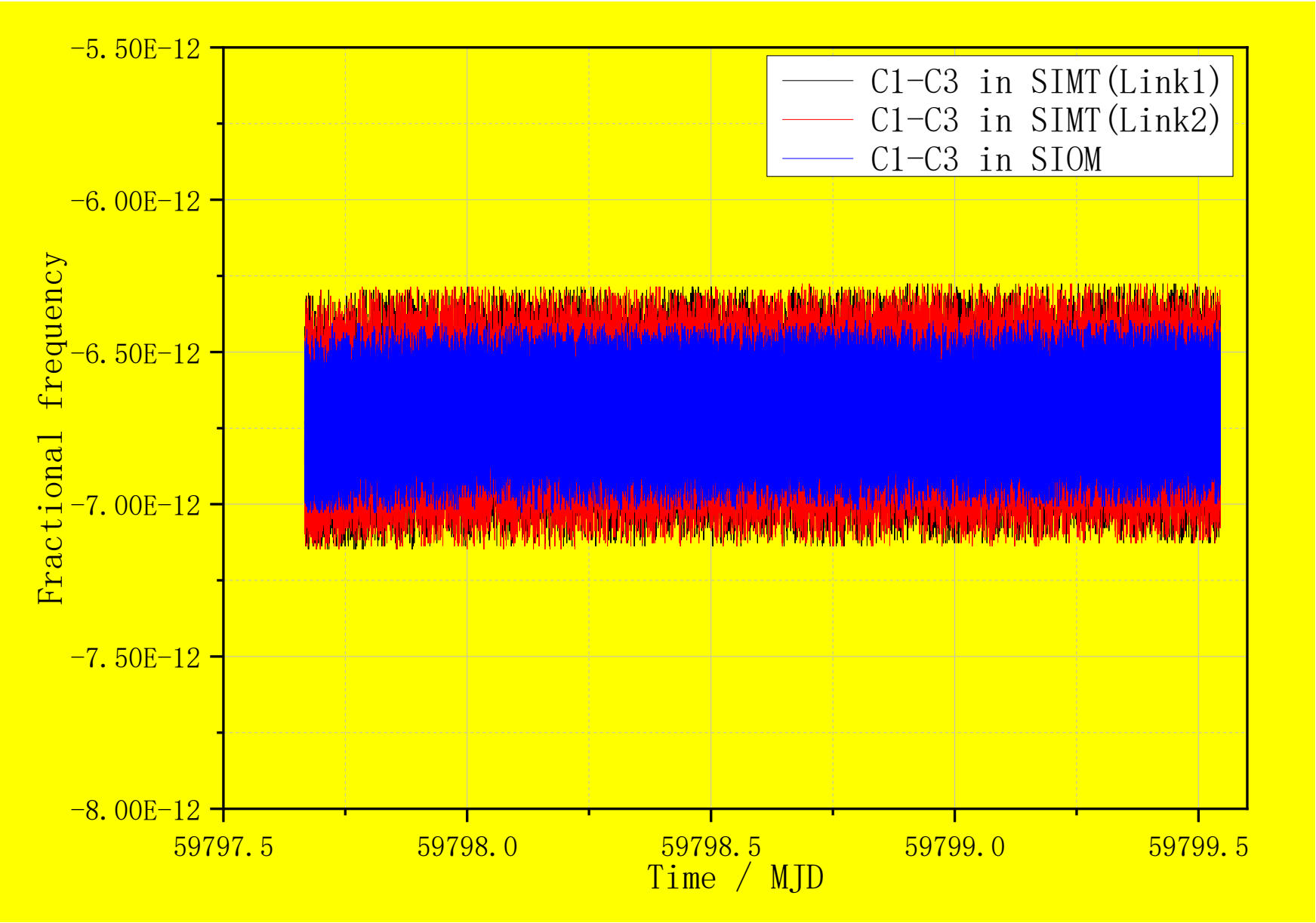

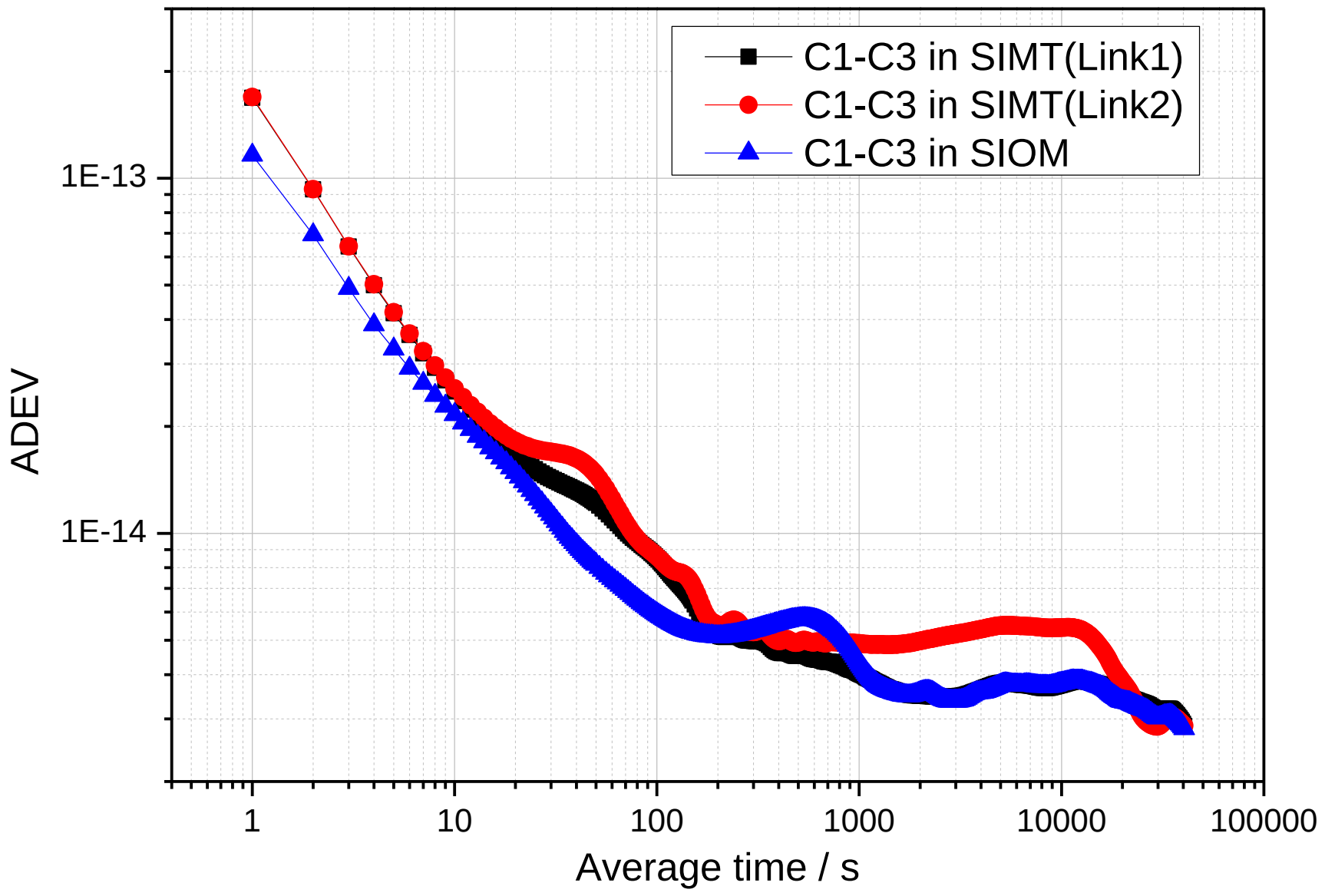

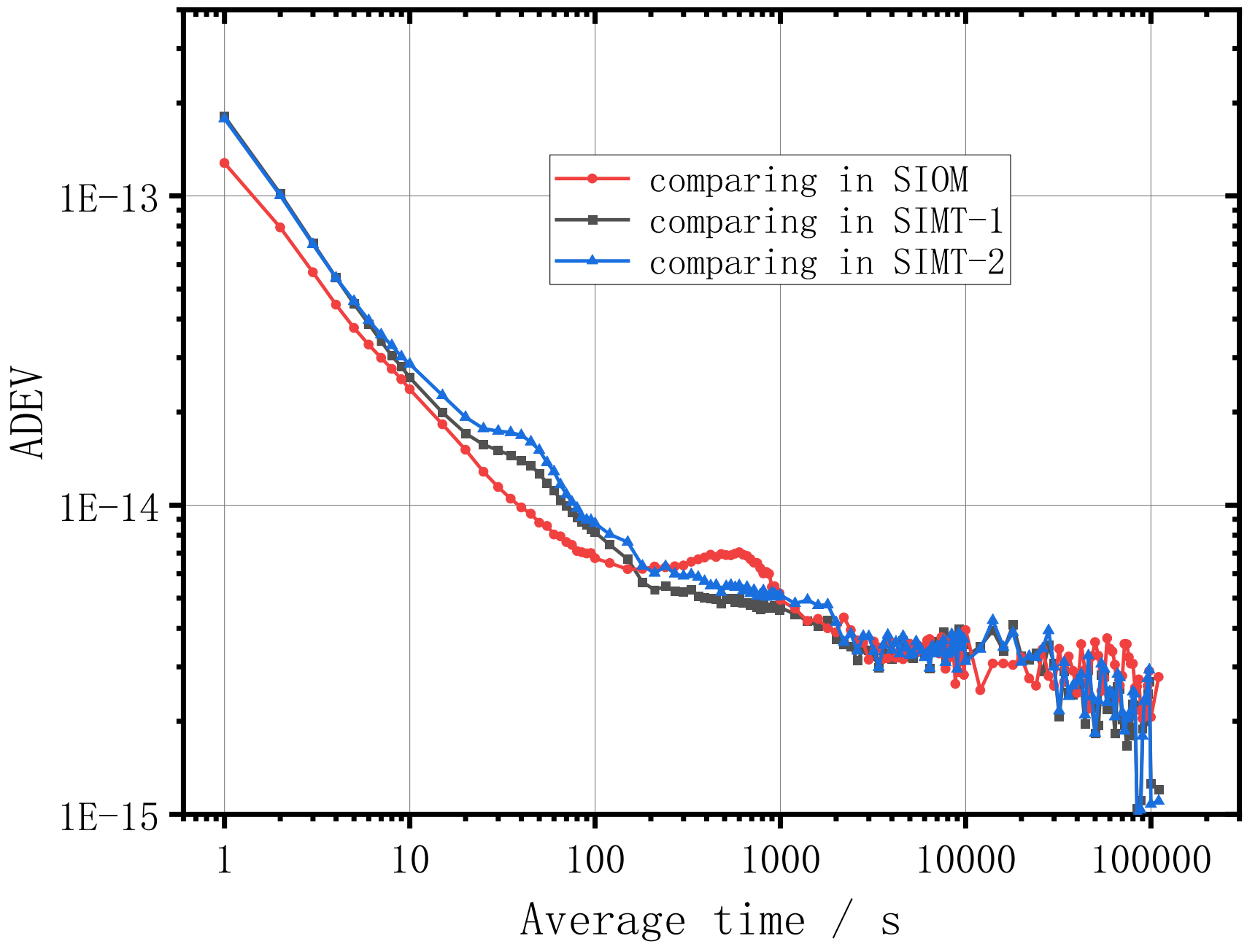

**Fig.3 Comparison results of CH.1 and CH.3 through 3 links**

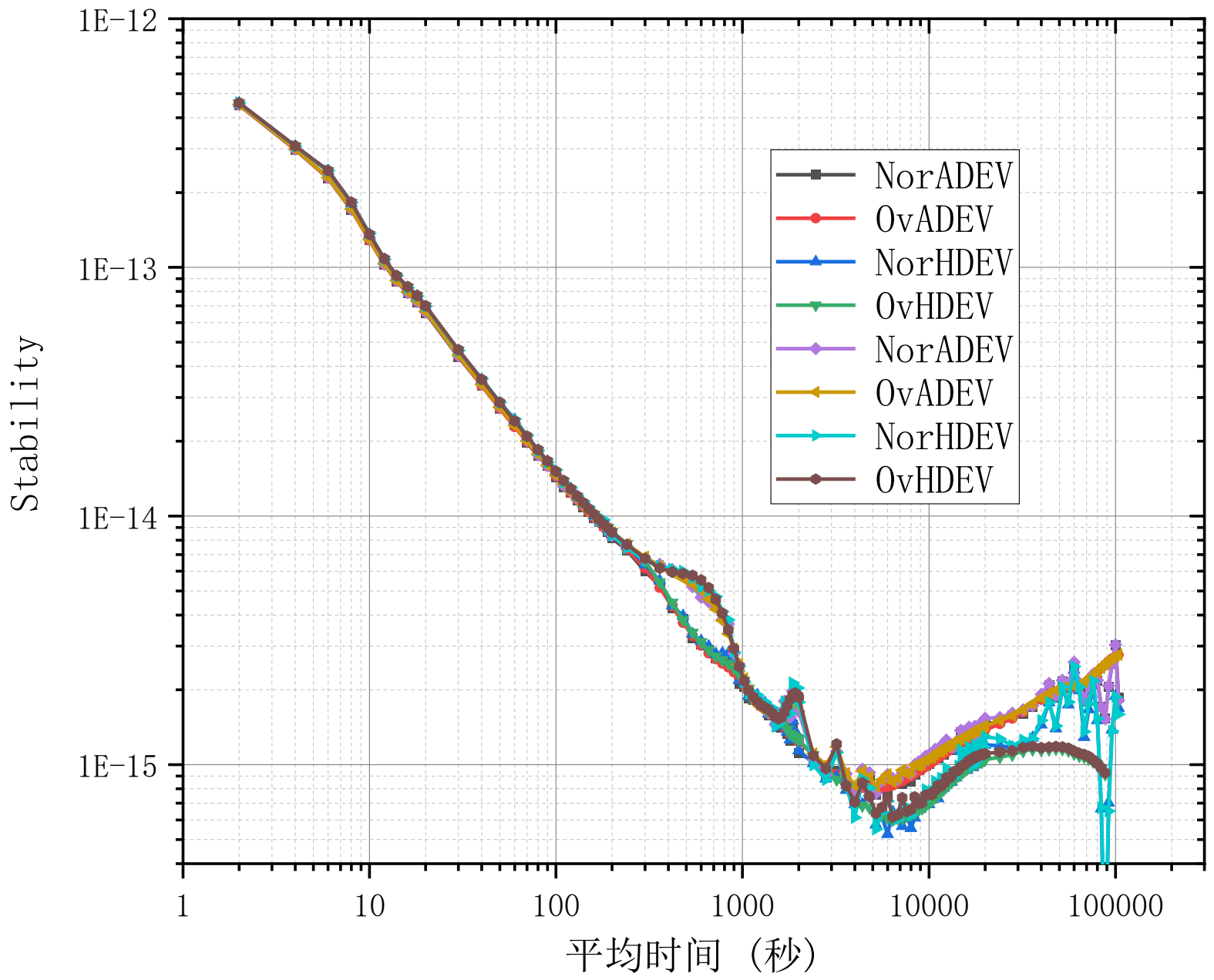

**Fig.4 Comparison results of two homologous signals inside SIOM**

In order to achieve the synchronous alignment of time-frequency signals between networks, we study a method of measuring the frequency difference of comparison data with a double-adoption correlation coefficient, and its basic principle is shown in the Appendix. According to this principle, we can use the method of plotting the correlation coefficient $R(y_1, y_2, l)$ curve, and then locate the maximum and minimum values of the curve to determine the relative time difference of the two data sets $L\tau_0$ and perform correction and data alignment. To verify the validity of the method, we first performed a test verification experiment at SIOM which is a comparison of C3 and C5 on two phase contrast instruments. The signal of C5 was passed through the 100 m fiber optic time frequency for inter-laboratory signaling, and the two signals were connected to two VCH314 phasers, and the comparison data were recorded by two computers. Since this is a relatively old set of fiber optic link, it uses a 100MHz modulation frequency, the performance index of short-term stability is relatively low, as shown in Figure 4, the short stability is around 6E-13@1s, and the noise mainly comes from the fiber optic link. We measured the standard covariance and double-sampling covariance with $R(y_{C3}, y_{C5}, l)$curves, where the double-sampling covariance is chosen for a total of 8 averaging times of 1s, 2s, ., $2s^7$.

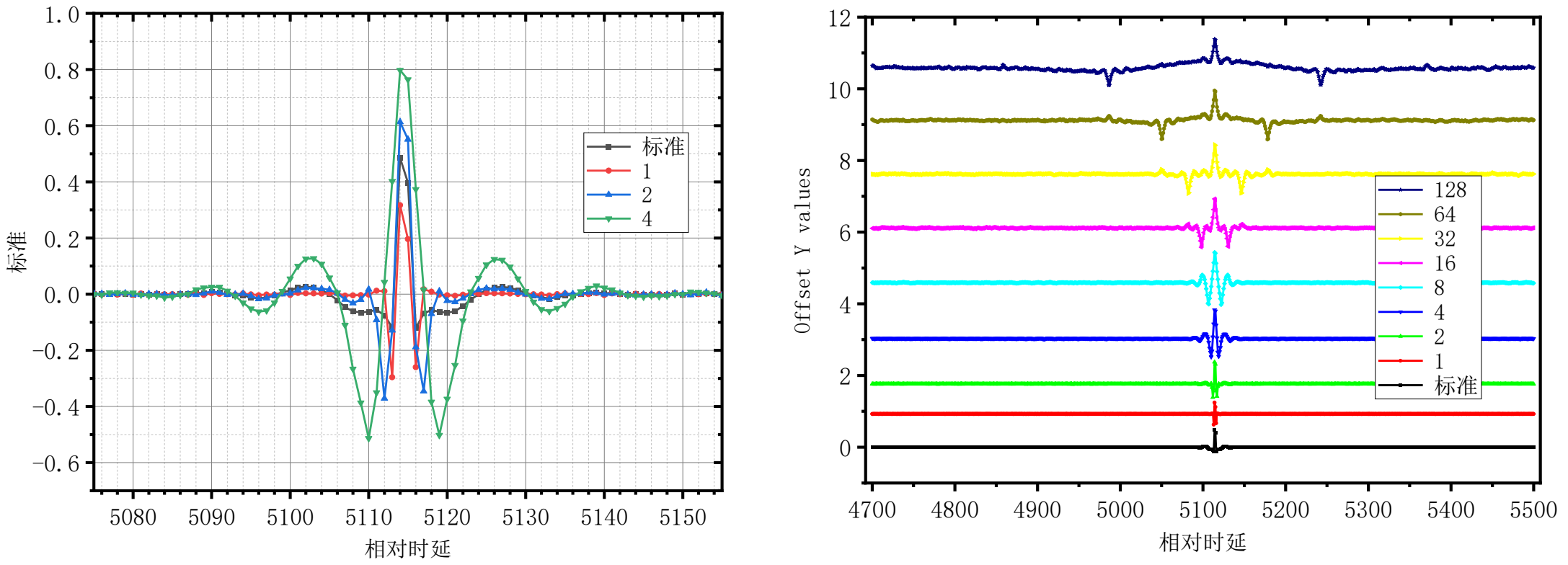

**Fig.5 Results of the comparison experiment**

The experimental results are shown in Fig. 5, and it can be seen that the same results are calculated with both the standard and double sampling methods, i.e., the maximum value corresponds to $L\tau_0 = 5114$s, and the theoretically expected double-sampling variance at $L\tau_0 \pm \tau$ position of the minimum value is also verified experimentally. lt can be seen fromthe figure that the $R(y_{C3}, y_{C5}, l)$ the actual extreme value should be between $L\tau_0 = 5114{\sim}5115$s between and closer to 5114s, which is about 5114.3s which means that the synchronization accuracy can be achieved below the sampling time.-We verified the clock synchronization error of the signal with the time created by the file and found that the two comparison data started with 8:42 and 10:06 on September 26, 2022 respectively, with a difference of 5040s. Considering that both computers that recorded the data did not have internet access and both had a deviation of several minutes, it can be assumed that the data from the measurement results are consistent with the actual results.

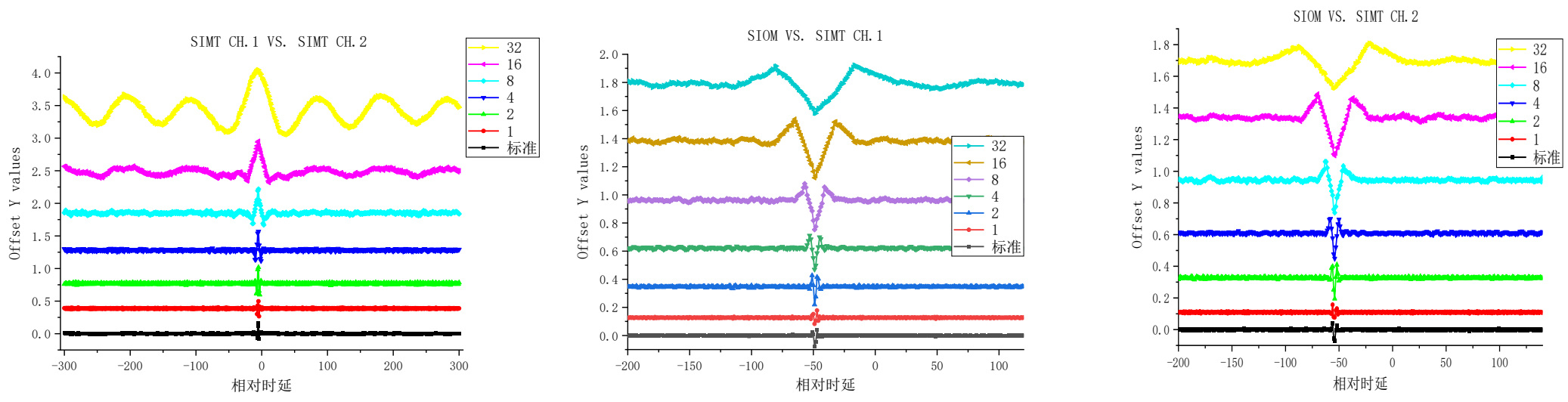

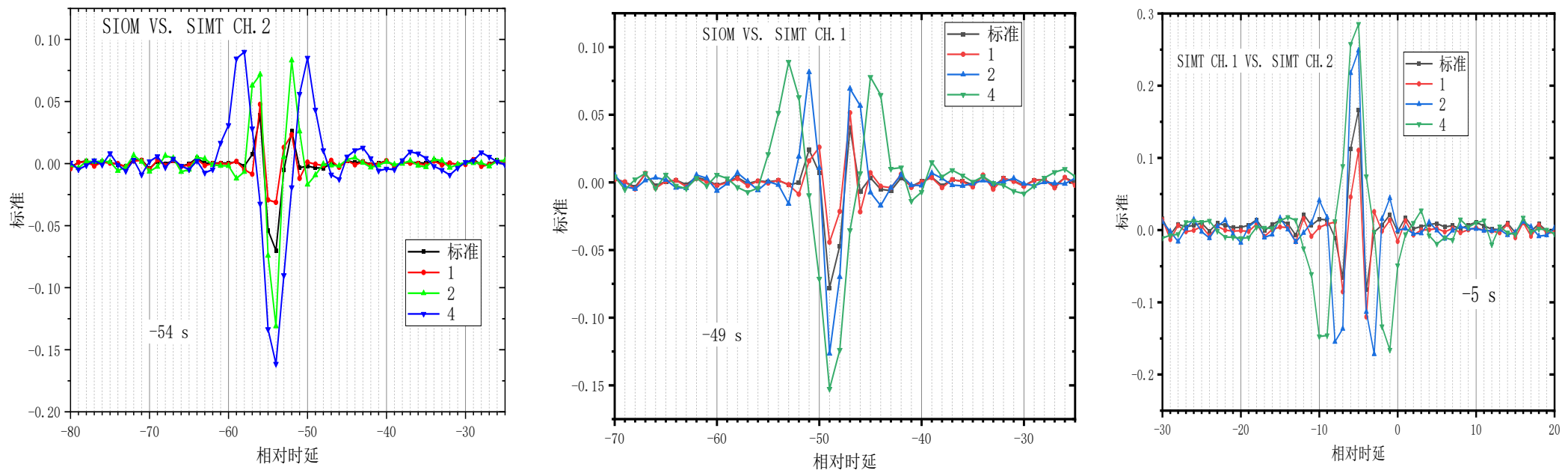

**Fig.6 Correlation results of the 3-link comparison data**

After completing the above verification experiment, we used the method to measure $\{y_{\mathrm{SIMT1}}\}$ and $\{y_{\mathrm{SIMT2}}\}$ and $\{y_{\mathrm{SIOM}}\}$. The correlation coefficient curves of the 3ratio noise sequences, and the experimental results are shown in Figure 6. lt can be seen that they are different from each other by 54s and 49s, and 5 s. The three times satisfy 54s and 49s and 5 s, and the data are consistent. Among them $\{y_{\mathrm{SIMT1}}\}$ and $\{y_{\mathrm{SIMT2}}\}$. The time difference of 5s between them is rather strange because they were recorded by the same phasor VCH 315.but the validity of the method was later verified by checking the data. Comparing Fig. 5 and Fig. 6, it can be seen that the great values of the R curves of the two laboratories are obviously much smaller than those of the SIOM local experiments, which is because the comparison experiments of the two laboratories contain more link noise, while the great values of the **R** curves in the SIOM local experiments are only affected by the microwave signal transmission and the comparison link.

We shipped the 3 sets of data differentially to see the impact of the correction for the time delay deviation. The difference between the two is equivalent to eliminating the bias of the comparison between the two clocks, and just measuring the relative bias of the two link comparisons it can be seen that the stability of the comparison signal has improved to different degrees after the correction of the time delay, although the degree of improvement is different, the $R(y_{\mathrm{C}i}, y_{\mathrm{C}j}, L)$ The greater the proportion of common mode noise in thepairwise comparison, the better the performance improvement of the linkwise comparison.This method can distinquish between atomic clock noise and link noise. locate the noise. And provide the basis for system improvement.

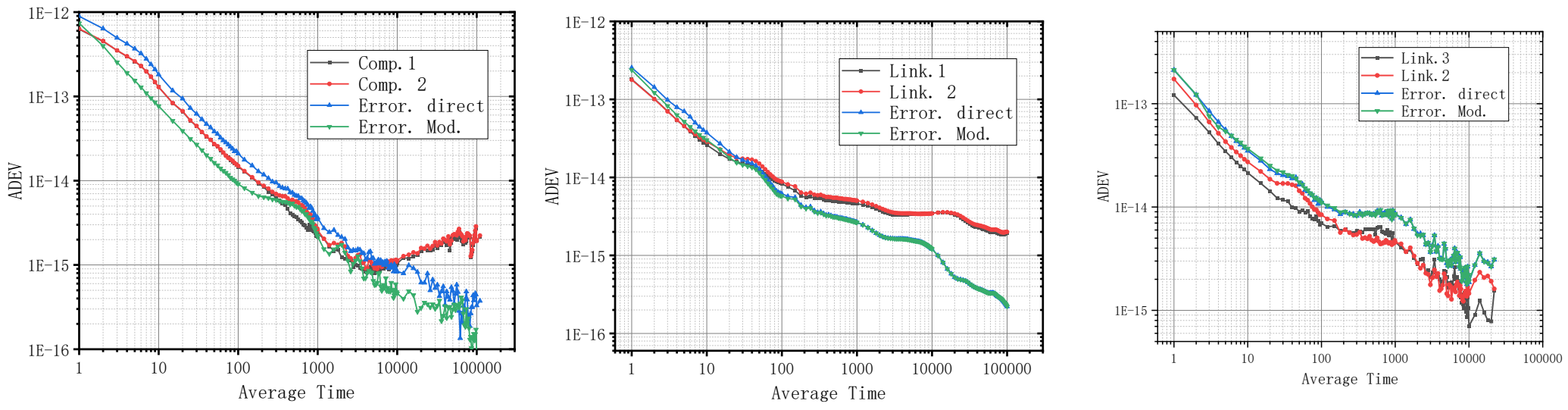

**Fig.7 Effect of error correction**

Using this method, we evaluated the time-frequency links of two distributed laboratories and realized the intercomparison of any two clocks from 5 atomic clocks in 2 laboratories.

## Appendix - Synchronization method of clock signals using correlation coefficients

It is often necessary to pass the clock signal remotely, corresponding to the scenario shown in Figure 7, with 3 clocks, which we denote by lowercase c0, c1 and c2 in order to distinguish them from the main text. c0 and cl, c0 and c2 are connected by links, but there is no link between c1 and c2 and now the comparison between c1 and c2 is calculated. We use $y_{\mathrm{c1}}$ and $y_{\mathrm{c2}}$ to denote the comparison results in two labs, Remote lab 1 and Remote lab 2. For simplicity, we count the noise of links fiber link 1 and 2 into c1 and c2, so that we have

$$y_{1,i}(\tau) = y_{\mathrm{c0},i}(\tau) - y_{\mathrm{c1},i}(\tau) \quad \text{(Attachment.1-1)}$$

$$y_{2,i}(\tau) = y_{\mathrm{c0},i}(\tau) - y_{\mathrm{c2},i}(\tau) \quad \text{(Attachment.1-2)}$$

Mathematically, the comparison result between c1 and c2 can be obtained by just the difference of the two formula $y_{\mathrm{c2},i}(\tau) - y_{\mathrm{c1},i}(\tau)$. However, this calculation is conditional and requires the two sets of comparison data to pe time-synchronized. in general, this condition is not satisfied, and it needs to be calculated by recording the data set $\{y_{1-i}(\tau)\}$ and $\{y_{2-i}(\tau)\}$ timestamp, synchronize the data according to the timestamp, and then perform the calculation. However, this also encounters some problems, because the timestamp needs to be traced back to standard time through the network and so on, and it has to be consistent with the data generation time of the ratio meter in hardware, while the computer recording the comparison data is often not connected to the network, and a long time running will produce errors of a few minutes or even longer. To solve this problem, we process the data with algorithms to obtain their time differences and correct them by finding correlations as away to achieve data alignment.

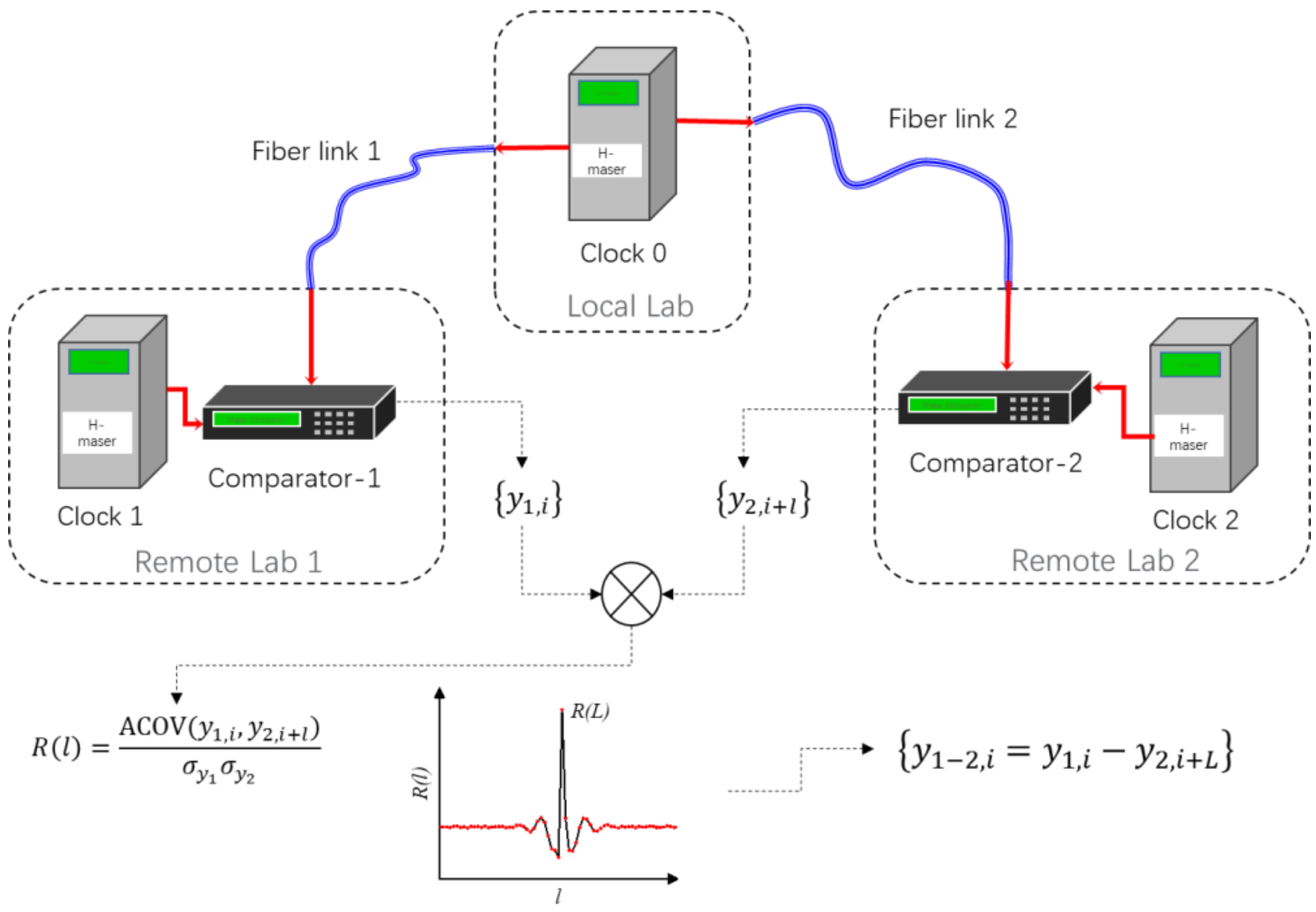

**Fig.8 Structure and schematic diagram of data synchronization by comparison**

The correlation between the two data sets is measured by the covariance and the correlation coefficient, and the correlation between the noises is assessed by the Autocovariance. To analyze the correlation, we introduce some new parameters by replacing the standard covariance with the double sampling covariance. Which is consistent with replacing the standard deviation with the Allan deviation for the time-frequency signal, the double sampling covariance is expressed as

$$\mathrm{ACOV}[y_1(\tau), y_2(\tau)] = \frac{1}{2(M-1)}\sum_{j=1}^{M-1}[y_{1,j+1}(\tau) - y_{1,j}(\tau)]\cdot[y_{2,j+1}(\tau) - y_{2,j}(\tau)] \qquad \text{(Attachment.2)}$$

the correlation coefficient for the time association is expressed as

$$R(y_1, y_2, l) = \frac{\mathrm{ACOV}(y_{1,i}, y_{2,i+l})}{\sigma_{y_1}\sigma_{y_2}} \qquad \text{(Attachment.3)}$$

of which $\sigma_{y_1}$ and $\sigma_{y_2}$ are the uncertainties obtained using the double sampling variancecalculation. Assuming that $\{y_{c0,i}(\tau)\}$ and $\{y_{c1,i}(\tau)\}$ and $\{y_{c2,i}(\tau)\}$ sequences areindependent of each other and their cross-multiplication terms sum to zero when there areenough elements of the time series, we have

$$\sigma_{y_1} \cong \sqrt{\sigma_{c0}^2 + \sigma_{c1}^2}, \quad \sigma_{y_1} \cong \sqrt{\sigma_{c0}^2 + \sigma_{c2}^2}, \qquad \text{(Attachment.4)}$$

$$\mathrm{ACOV}(y_{1,i}, y_{2,i+l}) \cong \mathrm{ACOV}(y_{c0,i}, y_{c0,i+l-L}) \quad \text{(Attachment.5)}$$

Here $L$ is the true asynchronous deviation of the two time series, the $L\tau_0$ is the time difference between the two data starts when $l = L$ when corresponds to $\{y_{c0,i}(\tau)\}$ the autocorrelation of $\mathrm{ACOV}(y_{c0,i}, y_{c0,i}) = \sigma_{c0}^2$ that is $\mathrm{ACOV}(y_{c0,i}, y_{c0,i+l-L})$ the maximum value of $R(y_1, y_2, l)$ the maximum value of $R(y_1, y_2, L)$. In this way, the relative time difference of the two data canbe determined by locating the maximum value $L\tau_0$ and thus data alignment can beperformed. And the $R(y_1, y_2, L)$ the magnitude of corresponds to the relative proportion of the noise in the stability of the c0 clock. And for the double adoption covariance, when $l = L \pm \tau/\tau_0$ When brought into equation (Attachment. 2). t can be seen that $\mathrm{ACOV}(y_{c0,i}, y_{c0,i+l-L})$ has a negative correlation minimum, the $R(y_1, y_2, L \pm \tau/\tau_0) \cong -R(y_1, y_2, L)/2$. As shown in the data in the main text, due to the influence of the correlation between the noises, we cannot obtain $R(y_1, y_2, l)$ the analytic equation of, and the actual $R(y_1, y_2, L \pm \tau/\tau_0)$ is also not fully satisfied, but the $R(y_1, y_2, L \pm \tau/\tau_0)$. The minimalvalue of is very clear, and we can further verify the value of according to the location of the minimal value $L$ The value of thee